\documentclass[prb,twocolumn,showpacs,preprintnumbers,amsmath,amssymb]{revtex4} 
\usepackage{graphicx}% Include figure files%

\begin{document}
 
\title{Electronic structure and magnetic anisotropy for 
nickel-based molecular magnets} 
\author{Kyungwha Park$^{1,2}$}\email{park@dave.nrl.navy.mil}
\author{En-Che Yang$^{3}$}
\author{David N. Hendrickson$^3$}
\affiliation{
$^1$Center for Computational Materials Science, Code 6390, 
Naval Research Laboratory, Washington DC 20375 \\
$^2$Department of Physics, Georgetown University, Washington DC 200057 \\
$^3$Department of Chemistry, University of California at San Diego,
La Jolla, CA 92093-0358} 
\date{\today} 
 
\begin{abstract}
Recent magnetic measurements on tetra-nickel 
molecular magnets [Ni(hmp)(ROH)Cl]$_4$, where R=CH$_3$, CH$_2$CH$_3$, 
or (CH$_2$)$_2$C(CH$_3$)$_3$ and hmp$^-$ is the monoanion of
2-hydroxymethylpyridine, revealed 
a strong exchange bias prior to the external magnetic field reversal
as well as anomalies in electron paramagnetic resonance peaks at low
temperatures. To understand the exchange bias and observed anomalies, 
we calculate the electronic structure and magnetic properties for 
the Ni$_4$ molecules with the three different ligands, employing
density-functional theory. Considering the optimized structure
with possible collinear spin configurations, we determine a
total spin of the lowest-energy state to be $S=0$, which
does not agree with experiment. We also calculate magnetic anisotropy 
barriers for all three types of Ni$_4$ molecules to be in the range 
of 4-6~K.
\end{abstract}

\maketitle

%\begin{multicols} 
%\section{Introduction}

Molecular magnets have received much attention in the past decade because 
of observed quantum features at the mesoscopic level and potential utilization 
as ultra-high density information storage devices or magnetic sensors. 
The first prototype molecular magnet is 
Mn$_{12}$O$_{12}$(CH$_3$COO)$_{16}$(H$_2$O)$_4$ (hereafter Mn$_{12}$-acetate),
\cite{LIS80} which has a total ground-state spin of $S=10$ and magnetic 
anisotropy barrier of 65~K.\cite{SESS93,FRIE96} 
Magnetic hysteresis loop measurements on Mn$_{12}$-acetate showed quantum 
tunneling of magnetic moment through the magnetic anisotropy barrier when 
the external magnetic field was applied along the easy axis.\cite{FRIE96} 
%The resonance magnetic fields where the tunneling occurred can be calculated
%using magnetic anisotropy parameters. 

\begin{figure}[h]
\includegraphics[angle=0,width=.23\textwidth,height=.27\textwidth]{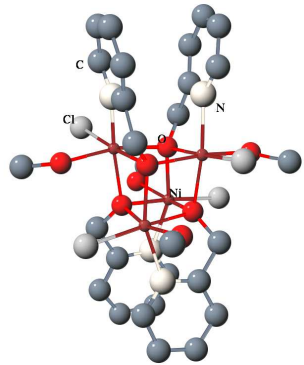}
\caption{Geometry of molecule magnet [Ni(hmp)(MeOH)Cl]$_4$. For simplicity,
hydrogen atoms are not shown.}
\label{fig:MeOH}
\end{figure}

Recently molecular magnets [Ni(hmp)(ROH)Cl]$_4$ (hereafter Ni$_4$) 
were synthesized,\cite{YANG03} where R=CH$_3$, CH$_2$CH$_3$, or (CH$_2$)$_2$C(CH$_3$)$_3$ 
(abbreviated to Me, Et, and tBuEt, respectively). 
They revealed qualitatively different tunneling features from the typical 
molecular magnets such as Mn$_{12}$-acetate. For Ni$_4$, the tunneling was 
observed prior to magnetic field reversal but no tunneling in zero magnetic 
field.\cite{YANG03} The similar behavior has been observed in a dimeric form of the 
molecular magnet [Mn$_4$O$_3$Cl$_4$(O$_2$CEt)$_3$(NC$_5$H$_5$)$_3$]$_2$ 
(hereafter Mn$_4$).\cite{HEND92,WERN02-NAT} Additionally, electron
paramagnetic resonance (EPR) spectra on Ni$_4$ exhibited unusual double sets of
low-temperature peaks corresponding to somewhat different values of magnetic 
anisotropy barriers.\cite{EDWA03} This anomaly in the EPR spectra has never been reported
from any other kinds of molecular magnets including the Mn$_4$ dimer. 
%The EPR peaks for the molecular magnets Ni$_4$ with MeOH, EtOH, and tBuEtOH 
%ligands were split into two types with slightly different values of the 
%magnetic anisotropy parameters with the same values of the Land\'{e} $g$ factor.
%The difference between the two types in the magnetic anisotropy parameters
%increased as smaller ligands in size were included in the molecular magnet Ni$_4$.
Each Ni$_4$ molecule comprises four Ni$^{2+}$ ($S_i=1$) ions located at the
corners of a slightly distorted cube and coupled through oxygen anions
with S$_4$ symmetry shown in Fig.~\ref{fig:MeOH}. 
Unlike Mn$_{12}$-acetate and Mn$_4$ dimer, the Ni$_4$ molecular magnets 
are not well separated from each other so that there are four
interacting nearest neighboring molecules.\cite{YANG03} The intermolecular 
interactions can be tailored by different sizes of ligands such as 
MeOH, EtOH, and tBuEtOH. Among the three different types of Ni$_4$, 
the MeOH and EtOH complexes consist of two sublattices, each of which is 
an inter-penetrating diamond lattice.\cite{YANG03}

In this paper, as an attempt to better understand the observed 
tunneling aspects and EPR spectra, preliminary first-principle 
calculations on the Ni$_4$ molecular magnets will be presented with 
the three different ligands using density-functional theory (DFT). 
Geometries of the Ni$_4$ molecular magnets with the different ligands 
will be relaxed and electronic density of states will be discussed
along with ground-state magnetic moments and magnetic anisotropy. 
All of our calculated results will be compared with available experimental 
data.

%\section{Density-Functional Calculations and Discussion}

\begin{table}
\begin{center}
\caption{Calculated energies (in units of eV) of different collinear spin 
configurations relative to the lowest-energy configuration ($M_s=0$) for 
the three types of Ni$_4$. Here MeOH.01 represents a molecule belonging to
the first sublattice of the MeOH complex.}
\label{table:Ms}
\begin{ruledtabular}
\begin{tabular}{|c|c|c|c|}
 & E($M_s=4$) & E($M_s=2$) & E($M_s=0$)   \\ \hline
MeOH.01 & 0.0308 & 0.0023 & 0.0000  \\ \hline
EtOH.01 & 0.0289 & 0.0042 & 0.0000  \\ \hline
tBuEtOH & 0.0283 & 0.0056 & 0.0000  \\ 
\end{tabular}
\end{ruledtabular}
\end{center}
\end{table}

Density-functional calculations\cite{KOHN65} were performed with spin-polarized
all-electron Gaussian-orbital basis sets and the generalized-gradient approximation
(GGA).\cite{PERD96} For this purpose, the Naval Research Laboratory Molecular Orbital 
Library (NRLMOL)\cite{PEDE90} is used. Compared to Mn$_{12}$-acetate and Mn$_4$ 
dimers, the Ni$_4$ molecular magnets need much finer mesh for faster convergence
of total energy.
Five different Ni$_4$ molecules are taken into account in density-functional
calculations: a tBuEtOH complex and two crystallographically inequivalent molecules 
of the MeOH and EtOH complexes. For each molecule, a relaxed geometry is 
found starting from a ferromagnetic $M_s=4$ spin configuration until forces 
exerted on any atom become relatively small (where $2M_s$ is magnetic moment). 
Although the moment is not fixed during self-consistent iterations, it converges 
to $M_s=4$. With the relaxed geometry, two other collinear spin configurations, 
$M_s=2$ and $M_s=0$, are considered as starting magnetic moments to calculate 
energies and compare to the energy of $M_s=4$. As shown in Table~\ref{table:Ms}, 
for the MeOH, EtOH, and tBuEtOH complexes, the $M_s=0$ configuration has the lowest
energy. This is in contrast to the experimental result\cite{YANG03,EDWA03}
which exhibited a ferromagnetic coupling $S=4$ as the ground state for all 
five Ni$_4$ molecules. Even though not shown in Table~\ref{table:Ms}, there 
are no differences in total moments of the ground states for the two 
inequivalent molecules of the MeOH and EtOH complexes. 
An experiment on a tetra-nickel-based molecular magnet within 
a Mo$_{12}$ cage, [Mo$_{12}$O$_{30}$($\mu_2$-OH)$_{10}$H$_2$\{Ni(H$_2$O)$_2$\}$_4$],
however, revealed that the ground state has a total spin of $S=0$.\cite{MULL00} 
The experimental 
finding of the ground-state spin was confirmed by density-functional calculations 
carried out by Postnikov {\it et al.}\cite{POST04} In this system, two Ni ions
are bridged via O-Mo$_2$-O, while in our system Ni ions are bridged via
oxygen anions only.  

\begin{figure}[h]
\includegraphics[angle=0,width=.4\textwidth,height=.3\textwidth]{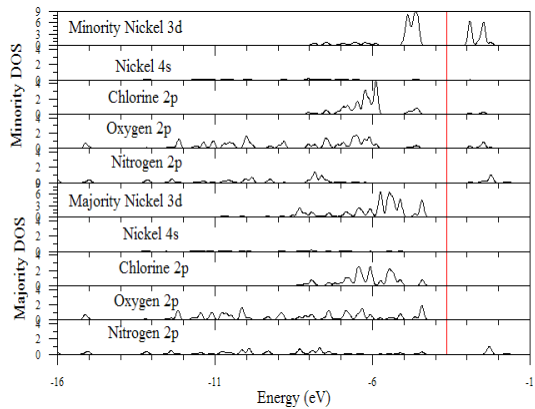}
\caption{Electronic majority- and minority-spin density of states (DOS) 
projected onto Ni(3d), Ni(4s), Cl(2p), O(2p), and N(2p) of 
[Ni(hmp)(EtOH)Cl]$_4$. All projected DOS have the same scale except
for Ni(3d) DOS. The vertical line denotes the Fermi level.}
\label{fig:DOS}
\end{figure}

\begin{table}
\begin{center}
\caption{Magnetic moments captured by spheres around ions in units
of $\mu_B$ for $M_s=4$ of the EtOH complex and for $M_s=10$ of the 
Mn$_{12}$-acetate. For Ni$_4$, O(b) denotes a bridging oxygen and 
O(l) represents a dangling oxygen from the cube. For Mn$_{12}$-acetate,
O(1) [O(2)] bridges Mn$^{4+}$(1) [Mn$^{3+}$(2)] ions. }
\label{table:moments}
\begin{ruledtabular}
\begin{tabular}{cccccc}
 Ni & Cl & O(b) & O(l) & N & \\ \hline
1.57 & 0.053 & 0.166 & 0.043 & 0.049 & \\ \hline \hline
 Mn$^{4+}$(1) & Mn$^{3+}$(2) & Mn$^{3+}$(3) & O(1) & O(2) & O(3) \\ \hline
 -2.58 & 3.62 & 3.55 & 0.012 & -0.024 & -0.029 \\
\end{tabular}
\end{ruledtabular}
\end{center}
\end{table}

To investigate what may cause the discrepancy between theory and experiment,
we calculate the magnetic moment captured by each atom, the electronic density of
states projected on individual atom, and the spin-flip energy gaps for the $M_s=4$
configuration. Within a sphere of 2.3 bohr radius around a Ni ion, a moment of 
1.6$\mu_B$ is found. The second dominant contribution to the total moment arises
from oxygen anions bridging Ni ions in the cube. A moment of 
0.17~$\mu_B$ is found within a sphere of 1.2 bohr radius around one of the oxygen anions. 
This is approximately 10\% of the captured moment around the Ni ion. Compared to 
a ferrimagnetic structure of the Mn$_{12}$-acetate and Mn$_4$ 
monomer,\cite{PEDE99,PARK04-Mn12,PARK03,PARK04-Mn4} somewhat 
larger moments are found around the oxygen anions in Ni$_4$ (Table~\ref{table:moments}).  
Interestingly, for the $M_s=0$ configuration, a bridging O anion 
in Ni$_4$ bears a moment of 0.073~$\mu_B$ within a sphere of 1.2 bohr radius.
From the calculated electronic density of states (Fig.~\ref{fig:DOS}) 
projected onto Ni 3d orbitals for $M_s=4$, we find that most of the electron 
spin density is localized at the Ni sites.
%Ni 3d orbitals for majority spin are all occupied, 
%while some of the minority-spin 3d orbitals are not occupied. 
Notice that some of the oxygen 2p orbitals for minority spin 
are not occupied, although all majority-spin O 2p orbitals are occupied.
This might imply a possible small leakage of spin density to the O sites. 
%Based on our preliminary investigation, it is speculated that magnetic moments 
%of the Ni ions might be less localized than those of the Mn ions. 
However, more scrutinized studies are required to conclude that this, in fact, 
attributes to the antiferromagnetic coupling in the calculated lowest-energy 
state. Another intriguing point is that as shown in Table~\ref{table:Ms},
collinear spin excitations for Ni$_4$ have an order 
of magnitude lower energies than those for the 
Mn$_{12}$-acetate\cite{PARK04-Mn12} and Mn$_4$ monomer\cite{PARK03,PARK04-Mn4}. 
A majority-minority spin-flip gap ranges from 1.4~eV to 1.6~eV and 
a minority-majority spin-flip gap varies between 1.8~eV and 2.1~eV 
depending on ligands as shown in Table~\ref{table:HL}. It seems that the $M_s=4$ 
configuration appears to be magnetically stable since the spin-flip gaps are much
larger than thermal energy. 
%The energy gaps slightly increase with increasing the volume of ligands.

\begin{table}
\begin{center}
\caption{Majority LUMO(lowest unoccupied molecular orbital)-
HOMO(highest occupied molecular orbital) gaps, minority 
LUMO-majority HOMO gaps, majority LUMO-minority HOMO gaps, 
minority HOMO-LUMO gaps in units of eV, and magnetic 
anisotropy barriers (MAB) in units of kelvin for $M_s=4$ of the 
three types of Ni$_4$.}
\label{table:HL}
\begin{ruledtabular}
\begin{tabular}{|c|c|c|c|c|c|}
 & Mj L-H & Mn L-Mj H & Mj L-Mn H &
Mn L-H & MAB \\ \hline
MeOH.01 & 1.74 & 1.38 & 1.83 & 1.47 & 4.4  \\ \hline
%MeOH.02 & 1.93 & 1.43 & 2.04 & 1.54 & 4.8 \\ \hline
EtOH.01 & 2.02 & 1.47 & 2.18 & 1.62 & 5.9 \\ \hline
%EtOH.02 & 2.10 & 1.54 & 2.23 & 1.67 & 4.0 \\ \hline
tBuEtOH & 1.97 & 1.57 & 2.12 & 1.71 & 5.4 \\
\end{tabular}
\end{ruledtabular}
\end{center}
\end{table}

An expectation value of the orbital angular momentum in the ground state of 
molecular magnets is typically quenched so that the ground-state spin manifold 
has a $(2S+1)$-fold degeneracy. The degeneracy can be lifted by spin-orbit coupling,
which leads to magnetic anisotropy in the system. A generalized second-order 
single-spin Hamiltonian is given by
\begin{equation}
{\cal H} = -D S_z^2 + E(S_x^2 - S_y^2)
\label{eq:ham}
\end{equation}
where $z$ is the easy axis, $D$ is the uniaxial anisotropy parameter, and 
$E$ is the transverse anisotropy parameter. To calculate the energy shift 
of the ground state, we consider the 
spin-orbit coupling as a small perturbation and use the computed single-electron
orbitals and orbital energies\cite{PEDE99} for the relaxed $M_s=4$ geometry. 
Because of S$_4$ symmetry of the Ni$_4$ molecule, the first term in Eq.~(\ref{eq:ham})
only survives. Calculated magnetic anisotropy barriers ($=DM_s^2$) for 
all five different Ni$_4$ molecules are in the range of 4-6~K so that
$D=0.25-0.37$~K. See Table~\ref{table:HL}. Measured EPR spectra on the five 
Ni$_4$ molecules revealed that values of $D$ are in the range of 0.72-1.03~K 
and that the two inequivalent molecules of the MeOH and EtOH complexes have 
considerably different values of $D$.\cite{EDWA03} The calculated values of
$D$ are merely 36\% of the experimentally extracted values.
Numerical uncertainties in density-functional calculations prevent from 
differentiating between the anisotropy barriers of the two inequivalent 
molecules of the MeOH and EtOH complexes or between the barriers of 
the Ni$_4$ molecules with different ligands. A single-ion anisotropy projected
\cite{BARU02,PARK04-Mn12}
onto a Ni ion in the MeOH complex provides local $D=1.52$~K and local $E=0.47$~K.
The local easy axis of the Ni ion is found to be tilted by 10.5$^{\circ}$ 
from the global easy axis. Due to lack of Jahn-Teller distortion, a small 
value of the local $D$ is theoretically expected for Ni$^{2+}$ ions.

%\section{Conclusion}

In summary, we have considered the Ni$_4$ molecular magnets with three
different ligands in volume and carried out first-principle calculations
of the electronic and magnetic structure. We found that 
the calculated lowest-energy state has a total spin of $S=0$, 
although experimental data indicated $S=4$. This is quite unusual 
because density-functional theory, so far, predicted/confirmed correct 
ground-state spins for a variety of molecular magnets such as the
Mn$_{12}$-acetate\cite{PEDE99,PARK04-Mn12}, Mn$_4$ monomer\cite{PARK03,PARK04-Mn4},
Co$_4$ molecule\cite{BARU02}, and Fe$_4$ molecule\cite{KORT03}. 
Although the reason has not yet been clarified, our preliminary studies 
tentatively suggested a plausible incomplete localization of spin density 
on Ni ions. There were no prominent effects of 
the ligands in different volume on the electronic structure and magnetic 
anisotropy of the Ni$_4$ molecules. 

%\begin{center}
%{\textbf{Acknowledgments}}
%\end{center}
K.P. is grateful to M. R. Pederson and S. Hill for helpful discussion. K.P. 
was supported in part by ONR and the DoD HPCMO CHSSI program. E.-C.Y. and D.N.H. 
were supported by NSF Grants Nos. CHE-0123603, CHE-0350615, and DMR-0103290.

%\clearpage

%\end{multicols} 

\end{document}